# Joint probability densities of an active particle coupled to two heat reservoirs

Jae Won Jung [a], Sung Kyu Seo [a], and Kyungsik Kim [a,b,*]

[a] *DigiQuay Ltd., Manan-Gu Anyang, Gyeonggido 14084, Republic of Korea*
[b] *Department of Physics, Pukyong National University, Busan 48513, Republic of Korea*

In this paper, we derive an altered Fokker-Planck equation for an active particle with the harmonic, viscous, and random forces, coupled to two heat reservoirs. We attain the solution for the joint distribution density of our topology, including the center topology, the ring topology, and the chain topology, subject to an exponential correlated Gaussian force. The mean squared displacement and the mean squared velocity behavior as the super-diffusions in $t \ll \tau$ and for $\tau = 0$, while those have the Gaussian forms in $t \gg \tau$ and for $\tau = 0$, where $\tau$ is the correlation time. We concomitantly calculate and analyze the non-equilibrium characteristics of the kurtosis, the correlation coefficient, and the moment from the derived moment equation.

--------------------------------------------------------------------------------------------------------------------------------

* Corresponding authors. E-mail: kskim@pknu.ac.kr

## 1. Introduction

The dynamical behavior for the passive and active particles in contact with the heat reservoir is easy to understand and handle, but currently it is a tough assignment [1,2,3]. As is well-known in statistical physics for various years, the motion of Brownian particles, which are small, somewhat larger than many surrounding particles, describes as the two forces, i.e. the fluctuating force caused by the action of the surrounding particles and viscous force resisted by the medium [4,5]. This motion is basically expressed in the Langevin equation, which has extended to the fluctuation-dissipation theorem. Currently, we can divide into the two-particle groups; the passive particle moving to no fixed position by the spatial interactions with other particles like the Brownian particle and the active particle moving with a sense of purpose like a micro-swimmer. At present, the dynamical behaviors for the micro-active and macro-passive particles are theoretically, numerically, and experimentally excavated, and are underway novel study. The dynamical behavior of particles has analyzed mainly by solving the generalized Langevin equation and the Fokker-Planck equation [6]. In addition, it is well known that the Fokker-Planck equation has variously studied in statistical physics, solid physics, quantum optics, applied statistics, chemical physics, theoretical biology, and circuit theory.

The energy flowing from the outside heat reservoir, the heat flow of transport and resistance, and the thermal and random noises describe the motion of the particles [7,8]. The force and the energy have deduced the dynamical behavior of particles through the moments of motion if we do not analyze and solve the probability distribution function. However, it is difficult to well describe the diffusion or the transport process by the Laplace transform or the Fourier transform of the probability distribution function having two variables, the displacement and the velocity. Via the diffusion and transport processes, the study of traps in the system of particles or colloids using harmonic potential has calculated and analyzed for the pastime [9,10], and such cases have been mainly numerically analyzed in the quantum mechanical open system [11,12] more recently than in the classical case.

For the passive and active particles, it is difficult to find the closed-form solutions for the joint distribution of position, direction, and velocity using the Fokker Planck equation, but the Laplace transform method has been used to find the spatial evolution of all dynamic moments in the arbitrary dimensions. In particular, as a function of activity and inertia, the steady state velocity distribution shows a remarkable reentrant crossover from passive Gaussian to active non-Gaussian behavior [9]. Some studies have constructed a corresponding phase diagram using an exact representation of the *d*-dimensional kurtosis, and such calculations have predicted to give as the mean squared displacement of the run-and-tumble particle and the active Ornstein-Uhlenbeck process [10]. The useful and challenging problems of treating the exact analytical properties of the Fokker-Planck equation and the generalized Langevin equation are more complicated and formidable. In efforts to surmount and resolve the theoretical and computer-simulating difficulties, the diffusion, transport, and advection models have investigated until now.

The microscopic and macroscopic motions for the passive and active particles with non-Markovian effect have recently attracted much interest in the physical systems. Examples of such systems are viscous suspending particles [1,13], time delayed systems [14-17], glass-forming liquids [18], active and biological matter [19-23], reaction kinetics, protein dynamics [24], neuro-physics, quantum optics [25]. Other scientists have recently investigated on dynamical correlation functions [26], oscillatory decay [27], molecular friction [28], anomalous diffusion [29,30], oscillating currents [31,32], spontaneous active matter [33], quantum phase transitions [34], and non-Markovian systems [35-37].

In this paper, we derive the Fokker-Planck equation for an active particle contacted with the two heat reservoirs. We use the Fourier transform to the joint probability density of the velocity and the displacement and solve directly the joint probability density. The organization of this paper is as follows. In section 2, we derive the Fokker-Planck equation in our model. We next compute approximately dynamic solution and its mean squared displacement and velocity in the limits of $t \ll \tau$, $t \gg \tau$ and for $\tau = 0$, where $\tau$ is the correlation time. In section 3, we calculate numerically the kurtosis, the correlation coefficient, and the moment related to the position and the velocity. We compare and analyze the all-to-all topology with the ring, chain, and center topologies. Finally, we provide and account for a conclusion summarizing our key findings in section 4.

## 2. Motion in the coupling of the system and the heat reservoir

In this section, we derive the Fokker-Planck equation for an active particle coupled to the two heat reservoirs. The heat reservoir $z_i$ with the coupling strength $k_i$ and $b_i$ is determined by the coupling constant $d_i$ for $i = 1, 2$. Our topology is analogous to the all-to-all topology introduced in Ref. [2]. Fig. 1(a) is the plot of our topology, and we settle the dynamical behavior of our topology by setting the coupling constants $a_1 = a_2 \equiv a$ and $d_1 = d_2 \equiv d$.

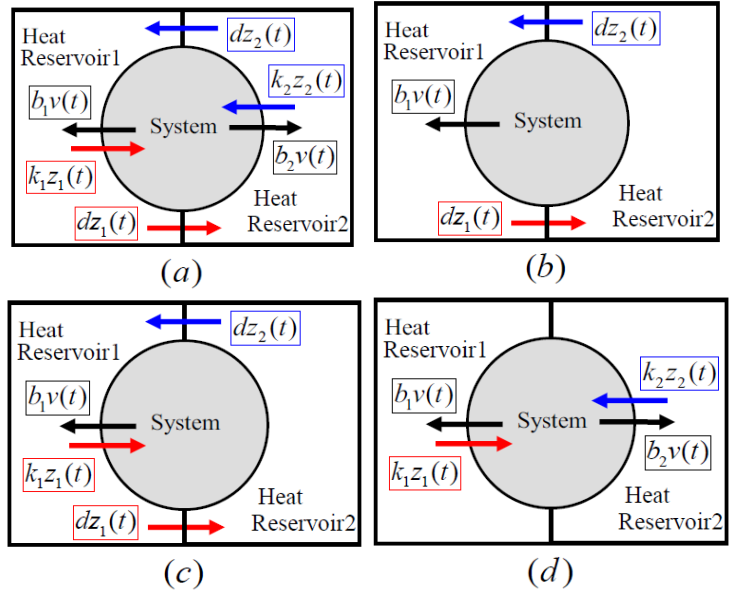

**Fig. 1.** Plots of (a) our topology, (b) the ring topology, (c) the chain topology, and (d) the center topology coupled to the two heat reservoirs 1 and 2.

We will solve the joint probability density for an active particle with two variables of the displacement and the velocity in the limits of $t \ll \tau$, $t \gg \tau$ and for $\tau = 0$, where $\tau$ is the correlation time.

### 2.1. The joint probability density $U(x, v, t)$

The following are the equations of motion for an active particle subjected to the harmonic and viscous forces, in contact with two heat reservoirs:

$$m_0 \frac{dv(t)}{dt} = -rv(t) - \beta x(t) + k_1 z_1(t) + k_2 z_2(t) + g_0(t), \tag{1}$$

$$m_1 \frac{dz_1(t)}{dt} = -a_1 z_1(t) + b_1 v(t) + d_1 z_2(t) + g_1(t), \tag{2}$$

$$m_2 \frac{dz_2(t)}{dt} = -a_2 z_2(t) + b_2 v(t) + d_2 z_1(t) + g_2(t). \tag{3}$$

Here, the zero-mean correlated Gaussian force $g_m(t)$ depends exponentially on the time difference: $<g_m(t)g_n(t')> = \frac{g_i^2}{2\tau}\exp(\frac{|t-t'|}{\tau})$. The force $-rv(t)$ denotes the viscous force, $-\beta x(t)$ the harmonic force, $m_i = 1$ the dimensionless mass, and $g_m^2 = k_B T_m \gamma_m \delta_{mn}$ ($m = 0,1,2$). For simplicity, treating only the case of $a_1 = a_2 = a$ and $d_1 = d_2 = d$ in order to get rid of the complicated property, we will attain the joint probability density in our model.

Next, the joint probability density $U(x(t),v(t),t)$ for the displacement $x$ and the velocity $v$ is defined by $U(x(t),v(t),t) = <\delta(x-x(t))\delta(v-v(t))>$. By taking time derivatives of the joint probability density $U(x(t),v(t),t) \equiv U(x,v,t)$, we have the differential equation as

$$\frac{\partial U(x,v,t)}{\partial t} = -\frac{\partial}{\partial x} < \frac{\partial x}{\partial t}\delta(x-x(t))\delta(v-v(t)) > -\frac{\partial}{\partial v} < \frac{\partial v}{\partial t}\delta(x-x(t))\delta(v-v(t)) > \cdot \quad (4)$$

By inserting Eq. (1) and Eq. (2) into Eq. (3), we directly take the Laplace transform in this equation. Using Eq. (4) and manipulating the method of Ref. [38], the Fokker-Plank equation for $U(x,v,t)$ with an exponentially Gaussian force is derived as

$$\frac{\partial}{\partial t}U(x,v,t) = [-v\frac{\partial}{\partial x} + (e+r)\frac{\partial}{\partial v}v + \beta x\frac{\partial}{\partial v}]U(x,v,t) + [-\alpha C(t)\frac{\partial^2}{\partial x \partial v} + \alpha B(t)\frac{\partial^2}{\partial v^2}]U(x,v,t) \cdot \quad (5)$$

Here, the parameters $e$, $\alpha$, $B(t)$, and $C(t)$ are, respectively, given by $e = -b_1 k_1 - b_2 k_2$, $\alpha = \alpha_0 + \alpha_1 k_1 + \alpha_2 k_2$, $B(t) = 1 - \exp(-t/\tau)$, and $C(t) = (t+\tau)\exp(-t/\tau) - \tau$, where $\alpha_i = g_i^2/2$ for $i = 0,1,2$. As a result, since the above Fokker-Planck equation does not include the two parameters $a$ and $d$, it is considered that these parameters do not affect the dynamical behavior of an active particle.

Now, the double Fourier transform of the joint probability density is defined by

$$P(\zeta,v,t) = \int_{-\infty}^{+\infty} dx \int_{-\infty}^{+\infty} dv \exp(-i\zeta x - ivv)P(x,v,t) \cdot \quad (6)$$

By using the double Fourier transform of Eq. (5), we derive the Fourier transform of the Fokker-Planck equation for an active particle as

$$\frac{\partial}{\partial t}U(\zeta,v,t) = [\zeta - (e+r)v]\frac{\partial}{\partial v}U(\zeta,v,t) - \beta v\frac{\partial}{\partial \zeta}U(\zeta,v,t) + \alpha[-B(t)v^2 + C(t)\zeta v]U(\zeta,v,t) \cdot \quad (7)$$

Later, we will derive the Fokker-Planck equation, Eq. (5), and the Fourier transform of the Fokker-Planck equation, Eq. (7), in Appendix A.

*2-2. $U(x,t)$ and $U(v,t)$ in the time domain $t \ll \tau$*

From Eq. (7), separating the variables as $U(\zeta,v,t) = U(\zeta,t)U(v,t)$, Eq. (7) becomes

$$\frac{\partial}{\partial t}U(\zeta,t) = [-\beta v\frac{\partial}{\partial \zeta} + \frac{\alpha}{2}[-B(t)v^2 + C(t)\zeta v] + k]U(\zeta,t), \quad (8)$$

$$\frac{\partial}{\partial t}U(v,t) = [\zeta - (e+r)v\frac{\partial}{\partial \zeta} + \frac{\alpha}{2}[-B(t)v^2 + C(t)\zeta v] - k]U(v,t) \cdot \quad (9)$$

Here, $k$ is the separation constant. Taking $\frac{\partial}{\partial t}U(\zeta,t) = 0$ for $\zeta$ in the steady state, $U(\zeta,t)$ is given by $U^{st}(\zeta,t)$. The Fourier transform for the joint probability density of $\zeta$ and $v$ is divided into the two equations by the variable-separation as

$$[-\beta v\frac{\partial}{\partial \zeta} + \frac{\alpha}{2}[-B(t)v^2 + C(t)\zeta v] + k]U^{st}(\zeta,t) = 0, \quad (10)$$

$$[\zeta - (e+r)v\frac{\partial}{\partial \zeta} + \frac{\alpha}{2}[-B(t)v^2 + C(t)\zeta v] - k]U^{st}(v,t) = 0 \cdot \quad (11)$$

The statistical value $U^{st}(\zeta,t)$ is calculated by

$$U^{st}(\zeta,t) = \exp[\frac{\alpha}{2\beta v}[\frac{C(t)}{2}v\zeta^2 - B(t)v^2\zeta]] \cdot \quad (12)$$

We now introduce a Fourier transform of the probability density $V(\zeta,t)$ by $U(\zeta,t) \equiv V(\zeta,t)U^{st}(\zeta,t)$. Using the above equation, then $U(\zeta,t)$ is derived by

$$U(\zeta,t) = V(\zeta,t)\exp[\frac{\alpha}{2\beta v}[\frac{C(t)}{2}v\zeta^2 - B(t)v^2\zeta] \cdot \quad (13)$$

By a similar method, we continuously derive that

$$V(\zeta,t) = W_\xi(\zeta,t)\exp[-\frac{\alpha}{2\beta v}[\frac{C'(t)}{6}v\zeta^3 - \frac{B'(t)}{2}v^2\zeta^2]], \quad (14)$$

$$W(\zeta,t) = Y(\zeta,t) \exp[\frac{\alpha}{2(\beta v)^3}[\frac{C''(t)}{24}v\zeta^4 - \frac{B''(t)}{6}]v^2\zeta^3], \qquad (15)$$

$$Y(\zeta,t) = Z(\zeta,t)\exp[-\frac{\alpha}{2(\beta v)^4}\frac{C'''(t)}{120}v\zeta^5]. \qquad (16)$$

Here, $B'(t) = dB(t)/dt$ and $C''(t) = d^2C(t)/dt^2$. As we neglect the terms proportional to $1/\tau^3$, $Z(\zeta,t)$ is given by $\frac{\partial}{\partial t}Z(\zeta,t) = -\beta v \frac{\partial}{\partial \xi}Z(\zeta,t)$. An arbitrary function of the variable $t-\zeta/\beta v$ becomes $Z(\zeta,t) = \Gamma[t-\zeta/\beta v]$. By expanding their derivatives to the second order in powers of $t/\tau$, we derive and calculate the expression for $U(\zeta,t)$ after some cancellations as

$$U(\zeta,t) = \Gamma[t-\zeta/\beta v]\, Y^{st}(\zeta,t)\, W^{st}(\zeta,t)\, V^{st}\zeta,t)\, U^{st}(\zeta,t)$$
$$= \exp[-\frac{\alpha t^3}{12\beta\tau}[1-\frac{3}{2}\frac{1}{\tau t}]\zeta^2 - \frac{\alpha t^3}{12\tau}[1-\frac{\tau}{6t}]v\zeta - \frac{\alpha t^3}{4\tau}[1+2\frac{\beta}{t}]v^2]. \qquad (17)$$

Using the inverse Fourier transform, the probability density $U(x,t)$ is presented by

$$U(x,t) = \frac{1}{2\pi}\int_{-\infty}^{+\infty} d\zeta \exp(-i\zeta x) U_\zeta(\zeta,t) = [\frac{\alpha t^3 \pi}{3\beta\tau}[1-\frac{3}{2\tau t}]]^{-1/2} \exp[-\frac{3\beta\tau x^2}{\alpha t^3}[1-\frac{3}{2\tau t}]^{-1}]. \qquad (18)$$

The mean squared displacement for $U(x,t)$ is given by

$$<x^2> = \frac{\alpha t^3}{6\beta\tau}[1-\frac{3}{2}\frac{1}{\tau t}]. \qquad (19)$$

In the short-time domain $t \ll \tau$, for obtaining the special solution for $v$, the first term of the left-handed side in Eq. (11) for $U(v,t)$ is assumed as $[\zeta-(e+r)v]^{-1} \cong \zeta^{-1}[1+(e+r)v/\zeta]$. The Fourier transform of the steady probability density $U^{st}(v,t)$ becomes

$$U^{st}(v,t) = \exp[\frac{\alpha}{2\zeta}[\frac{B(t)}{3}v^3 - \frac{C(t)}{2}\zeta v^2] + \frac{\alpha(e+r)}{2\zeta^2}[\frac{B(t)}{4}v^4 - \frac{C(t)}{3}\zeta v^3] + \frac{k}{\zeta}v + \frac{kr}{\zeta^2}\frac{v^2}{2}]. \qquad (20)$$

Similarly, it is expedient to perform the sequence of successive transformations as

$$U(v,t) = V(v,t)\exp[\frac{\alpha}{2\zeta}[\frac{B(t)}{3}v^3 - \frac{C(t)}{2}\zeta v^2] + \frac{\alpha(e+r)}{2\zeta^2}[\frac{B(t)}{4}v^4 - \frac{C(t)}{3}\zeta v^3]], \qquad (21)$$

$$V(v,t) = W(v,t)\exp[\frac{\alpha}{2\zeta^2}[\frac{B'(t)}{12}v^4 - \frac{C'(t)}{6}\zeta v^3] + \frac{\alpha(e+r)}{2\zeta^2}[\frac{B'(t)}{4}v^5 - \frac{C'(t)}{3}\zeta v^4]], \qquad (22)$$

$$W(v,t) = Y(v,t)\exp[\frac{\alpha}{2\zeta^3}[\frac{B''(t)}{60}v^5 - \frac{C''(t)}{24}\zeta v^4] + \frac{\alpha(e+r)}{2\zeta^4}[\frac{B''(t)}{120}v^6 - \frac{C''(t)}{60}\zeta v^6]], \qquad (23)$$

$$Y(v,t) = Z(v,t)\exp[-\frac{\alpha}{\zeta^3}\frac{C'''(t)}{240}v^5 - \frac{\alpha(e+r)}{\zeta^4}\frac{C'''(t)}{720}v^6]. \qquad (24)$$

Discarding the terms proportional to $1/\tau^3$, $Z(v,t)$ obeys

$$\frac{\partial}{\partial t}Z(v,t) = [\zeta-(e+r)v]\frac{\partial}{\partial v}Z(v,t). \qquad (25)$$

In Eq. (25), we satisfy the solution as a function of variable $t+v/[\zeta-(e+r)v]$, and an arbitrary function is given by $Z(\zeta,t) = \Gamma[t+v/[\zeta-(e+r)v]]$. Expanding their derivatives to the second order in powers of $t/\tau$, we derive and calculate the expression for $U(v,t)$ after some cancellations as

$$U(v,t) = \Gamma[t+\frac{v}{\zeta-(e+r)v}]Y^{st}(v,t)W^{st}(v,t)V^{st}(v,t)U^{st}(v,t) = \exp[-\frac{3\alpha(e+r)^2 t^3}{4}v^2 - \frac{\alpha t^4}{4\tau}[1-\frac{\tau}{t}]\zeta^2]. \qquad (26)$$

The inverse Fourier transform of $U(v,t)$ for the velocity $v$ yields

$$U(v,t) = \frac{1}{2\pi}\int_{-\infty}^{+\infty} dv \exp(-ivv)\, U(v,t) = [\pi 3\alpha(e+r)^2 t^3]^{-1/2} \exp[-\frac{v^2}{3\alpha(e+r)^2 t^3}] \qquad (27)$$

with its mean squared velocity for $U(v,t)$

$$<v^2(t)> = \frac{3\alpha(e+r)^2 t^3}{4}[1+\frac{2}{3(e+r)t}]. \qquad (28)$$

*2-3. $U(x,t)$ and $U(v,t)$ in the time domain $t \gg \tau$*

In this subsection, we focus on the probability densities $U(x,t)$ and $U(v,t)$ in the long-time domain. We can write approximate equation for *x* from Eq. (8) like

$$\frac{\partial}{\partial t}U_\zeta(\zeta,t) \cong \frac{\alpha}{2}[C(t)\zeta v - B(t)v^2]U_\zeta(\zeta,t). \qquad (29)$$

From the above equation, we briefly have

$$U_\zeta(\zeta,t) = \exp[\frac{\alpha}{2}\int [C(t)\zeta v - B(t)v^2]\,dt]. \tag{30}$$

We find $V_\zeta^{st}(\zeta,t)$ for $\xi$ from $U(\zeta,t) \equiv V_\zeta(\zeta,t)U_\zeta(\zeta,t)$ as

$$V_\zeta^{st}(\zeta,t) = \exp[-\frac{\alpha}{2}\int [C(t)\zeta v]\,dt]. \tag{31}$$

From Eq. (10), we calculate $U^{st}(\zeta,t)$ as

$$U^{st}(\zeta,t) = \exp[\frac{\alpha}{2\beta v}[\frac{C(t)}{2}v\zeta^2 - B(t)v^2\zeta]], \tag{32}$$

$$V(\zeta,t) = W(\zeta,t)V^{st}(\zeta,t) = W(\zeta,t)\exp[-\frac{\alpha}{2}\int [C(t)\zeta \eta]\,dt]. \tag{33}$$

Taking the solutions as arbitrary functions of variable $t - \zeta/\beta v$, the arbitrary function $W(\zeta,t)$ becomes $W(\zeta,t) = \Gamma[t - [\zeta/\beta v]]$. By expanding in powers of $t/\tau$, we derive and calculate the expression for $U(\zeta,t)$ after some cancellations, as

$$U(\zeta,t) = W(\zeta,t)V^{st}(\zeta,t)U^{st}(\zeta,t) = \Gamma[t - [\zeta/\beta v]]V^{st}(\zeta,t)U^{st}(\zeta,t) = \exp[-\frac{\alpha t}{2}[1+\tau]\zeta^2 + \frac{3\alpha\beta t^2}{4}v\zeta - \frac{\alpha\beta^2 t^3}{6}[1-\frac{3\tau}{\beta^2 t}]v^2]. \tag{34}$$

Here,

$$\Gamma(u) = \exp[\frac{a}{2}\beta\tau tv^2 u - \frac{a}{2}(t-\tau)v^2 + \frac{a\beta\tau}{4}v^2 u^2 - \frac{a}{2}v^2 u]. \tag{35}$$

In the long-time domain, we write the approximate equation for $v$

$$\frac{\partial}{\partial t}U_v(v,t) \cong \frac{\alpha}{2}[C(t)\zeta v - B(t)v^2]U_v(v,t). \tag{36}$$

The probability density $U_v(v,t)$ can be calculated as

$$U_v(v,t) = \exp[\frac{\alpha}{2}\int [C(t)\zeta v - B(t)v^2]\,dt]. \tag{37}$$

As we have $\int B(t)dt = t-\tau$, $B(t)=1$ and $\int C(t)dt = -\tau t$, $C(t) = -\tau$ in the long-time domain, we find the solution of functions for $v$ from $U(\zeta,t) = V(\zeta,t)U^{st}(\zeta,t)$ of Eq. (13) as

$$V^{st}(v,t) = \exp[\frac{\alpha}{2}\int [B(t)v^2 - C(t)\zeta v]\,dt]. \tag{38}$$

Now, we have for the long-time domain $t \gg \tau$

$$U^{st}(v,t) = \exp[\frac{a}{2\zeta}\int [1 + \frac{(e+r)v}{\zeta}][b(t)v^2 - \frac{c(t)}{2}\zeta v + A]\,dv]. \tag{39}$$

Taking the solutions as arbitrary functions of variable $t + v/[\zeta - rv]$, the arbitrary function $W(v,t)$ becomes $W(v,t) = \Gamma[t + v/[\zeta - (e+r)v]]$. By expanding in powers of $t/\tau$, we obtain the expression for $U(v,t)$ after some cancellations as

$$U(v,t) = W(v,t)V^{st}(v,t)U^{st}(v,t) = \Gamma[t + v/[\zeta - (e+r)v]]V^{st}(v,t)U^{st}(v,t). \tag{40}$$

Therefore, from Eq. (34) and Eq. (40), we calculate that

$$U(\zeta,v) = U(\zeta,t)U(v,t) = \exp[-\alpha(e+r)^2 t^3[1 + \frac{\tau}{2(e+r)^2 t}]v^2 - \frac{\alpha(e+r)^2 t^4}{2}[1 + \frac{\tau}{(e+r)^2 t}]v\zeta - \frac{\alpha t^4}{6}\zeta^2]. \tag{41}$$

From the above equation, using the inverse Fourier transform, the probability densities $U(x,t)$ and $U(v,t)$ are, respectively, presented by

$$U(x,t) = [2\pi\alpha t[1+\tau]]^{-1/2}\exp[-\frac{x^2}{2\alpha t}[1+\tau]^{-1}], \tag{42}$$

$$U(v,t) = [4\pi\alpha(e+r)^2 t^3[1 + \frac{\tau}{2(e+r)^2 t}]]^{-1/2}\exp[-\frac{x^2}{4\alpha(e+r)^2 t^3}[1 + \frac{\tau}{2(e+r)^2 t}]^{-1}]. \tag{43}$$

The mean squared displacement and the mean squared velocity for $P(x,t)$ and $P(v,t)$ are, respectively, given by

$$<x^2(t)> = \alpha t[1+\tau], \tag{44}$$

$$<v^2(t)> = 2\alpha(e+r)^2 t^3[1 + \frac{\tau}{2(e+r)^2 t}]. \tag{45}$$

2-4. $U(x,t)$ and $U(v,t)$ for $\tau = 0$

For $\tau = 0$ ($B(t)=1, C(t)=0$), we write the approximate equation from Eq. (8) and Eq. (9) for $\zeta$ and $v$ as

$$\frac{\partial}{\partial t}U(\zeta,t) = -\beta v \frac{\partial}{\partial \zeta}U(\zeta,t) - \frac{1}{2}\alpha v^2 U(\zeta,t), \tag{46}$$

$$\frac{\partial}{\partial t}U(v,t) = [\zeta - (e+r)v]\frac{\partial}{\partial v}U(v,t) - \frac{1}{2}\alpha v^2 U(v,t). \tag{47}$$

From the above equations in the steady state, we calculate $U^{st}(\zeta,t)$ and $U^{st}(v,t)$ as

$$U^{st}(\zeta,t) = \exp[-\frac{\alpha}{2\beta v}v^2 \zeta], \tag{48}$$

$$U^{st}(v,t) = \exp[\frac{\alpha}{2\zeta}\frac{v^3}{3} + \frac{\alpha r}{2\zeta^2}\frac{v^4}{4}]. \tag{49}$$

Here, we assume that $[\zeta-(e+r)v]^{-1} \cong \frac{1}{\zeta}[1+\frac{(e+r)v}{\zeta}]$. The Fourier transforms of the probability densities, $U(\zeta,t)$ and $U(v,t)$, are derived as

$$U(\zeta,t) = \Gamma[t - \frac{\zeta}{\beta v}]U^{st}(\zeta,t), \tag{50}$$

$$U(v,t) = \Gamma[t + v/[\zeta-(e+r)v]]U^{st}(v,t). \tag{51}$$

We calculate $U(\zeta,v,t)$ from Eq. (50) and Eq. (51) as

$$U(\zeta,v,t) = U(\zeta,t)U(v,t) = \exp[-\frac{\alpha(e+r)t^4}{8}[1+\frac{4}{3}[(e+r)t]^{-1}]\xi^2 - \frac{\alpha(e+r)t^3}{2}[1+[(e+r)t]^{-1}]\xi v - \frac{\alpha t}{2}[1-\frac{1}{4}[(e+r)t]^{-1}]v^2]. \tag{52}$$

By using the inverse Fourier transforms, $U(x,t)$ and $U(v,t)$ are, respectively, presented by

$$U(x,t) = [\pi \frac{\alpha(e+r)t^4}{2}]^{-1/2}\exp[-\frac{2x^2}{\alpha(e+r)t^4}], \tag{53}$$

$$U(v,t) = [2\pi\alpha t]^{-1/2}\exp[-\frac{v^2}{2\alpha t}]. \tag{54}$$

The mean-squared deviations for $U(x,t)$ and $U(v,t)$ are, respectively, calculated like

$$<x^2(t)> = \frac{\alpha(e+r)t^4}{4}[1+\frac{4}{3}[(e+r)t]^{-1}], \tag{55}$$

$$<v^2(t)> = \alpha t[1-\frac{1}{4}(e+r)t]. \tag{56}$$

## 3. Moment equation and four topologies

Multiplying and integrating both sides of the Fokker-Plank equation (Eq. (5)) by $x^m v^n$, we get the moment equation for $\mu_{m,n}$ as

$$\frac{d\mu_{m,n}}{dt} = m\mu_{m-1,n+1} - (e+r)n\mu_{m,n} + \beta n\mu_{m+1,n-1} - mnac(t)\mu_{m-1,n-1} + n(n-1)ab(t)\mu_{m,n-2}. \tag{57}$$

Here, $\mu_{m,n} = \int_{-\infty}^{+\infty}dx\int_{-\infty}^{+\infty}dv x^m v^n P(x,v,t)$.

The kurtosis for $x$ is given by

$$K_x = <x^4>/3<x^2>^2 - 1. \tag{58}$$

We lastly calculate the correlation coefficient as

$$\rho_{xv} = <(x-\bar{x})(v-\bar{v})>/\sigma_x \sigma_v. \tag{59}$$

Here, $\bar{x}, \bar{v}$ denote the mean displacement and mean velocity of the joint probability density, and $\sigma_x, \sigma_v$ denote the root-mean-squared displacement and the root-mean-squared velocity of the joint probability density, respectively. An active particle with radial and tangential forces is initially at $x = x_0$ and at $v = v_0$.

**Table 1.** Values of the Kurtosis, the correlation coefficient, and the moment for an active particle of our topology.

| Time | Variables | $K_x, K_v$ | $\rho_{xv}$ | $\mu_{2,2}$ |
|---|---|---|---|---|
| $t \ll \tau$ | $x$ | $\frac{\beta^2\tau^4 x_0^4}{\alpha^2}t^{-6} - \frac{\beta\tau^2 x_0^2}{\alpha}t^{-3}$ | $\frac{\beta^{1/2}\tau^{3/2}x_0 v_0}{\alpha}t^{-5/2}$ | $\frac{\alpha^2}{\beta\tau^3[1+2(e+r)t]}t^5$ |
| | $v$ | $\frac{\tau^2 v_0^4}{\alpha^2}t^{-4} - \frac{v_0^2}{\alpha}t^{-2}$ | | |
| $t \gg \tau$ | $x$ | $\frac{x_0^4}{\alpha^2(e+r)^2\tau^2}t^{-6} - \frac{x_0^2}{\alpha(e+r)\tau}t^{-3}$ | $\frac{x_0 v_0}{\alpha(e+r)^{1/2}\tau^{1/2}}t^{-2}$ | $\frac{\alpha^2 r\tau}{6[1+2(e+r)t]}t^4$ |
| | $v$ | $\frac{v_0^4}{\alpha^2}t^{-2} - \frac{v_0^2}{\alpha}t^{-1}$ | | |
| $\tau = 0$ | $x$ | $\frac{x_0^4}{\alpha^2(e+r)^2}t^{-8} - \frac{x_0^2}{\alpha(e+r)}t^{-4}$ | $\frac{x_0 v_0}{\alpha(e+r)^{1/2}}t^{-5/2}$ | $\frac{\alpha^2 r}{[1+2(e+r)t]}t^5$ |
| | $v$ | $\frac{v_0^4}{\alpha^2}t^{-2} - \frac{v_0^2}{\alpha}t^{-1}$ | | |

Next, the joint probability density $U(x,v,t)$ for an active particle with the harmonic and viscous forces in section 2 was derived as Eq. (5),

$$\frac{\partial}{\partial t}U(x,v,t) = [-v\frac{\partial}{\partial x} + (e+r)\frac{\partial}{\partial v}v + \beta x\frac{\partial}{\partial v}]U(x,v,t) + [-\alpha C(t)\frac{\partial^2}{\partial x\partial v} + \alpha B(t)\frac{\partial^2}{\partial v^2}]U(x,v,t). \tag{5}$$

Let us introduce and calculate briefly four coupling topologies from Eq. (5), that is, the all-to-all topology, the ring topology, the chain topology, and the center topology. The coupling constants is provided in Table 2, and Table 3 is also summarized the values of the joint probability density, the mean squared displacement and the mean squared velocity for our topology, the ring topology, the center topology, and the chain topology in the limits of $t \ll \tau$, $t \gg \tau$ and for $\tau = 0$, where $\tau$ is the correlation time.

**Table 2.** Values of the coupling constants $e$, $\alpha$, $a_1$, $a_2$, $b_1$, $b_2$, $d_1$, $d_2$, $k_1$, and $k_2$. in our topology, the ring topology, the chain topology, the center topology.

| Topology | Coupling constants | | | | | | | | | |
|---|---|---|---|---|---|---|---|---|---|---|
| | $a_1$ | $a_2$ | $b_1$ | $b_2$ | $d_1$ | $d_2$ | $k_1$ | $k_2$ | $e = -(k_1 b_1 + k_2 b_2)$ | $\alpha = k_1\alpha_1 + k_2\alpha_2 + \alpha_0$ |
| our topology | $a$ | $a$ | $b_1$ | $b_2$ | $d$ | $d$ | $k_1$ | $k_2$ | $-k_1 b_1 - k_2 b_2$ | $k_1\alpha_1 + k_2\alpha_2 + \alpha_0$ |
| ring | $a$ | $a$ | $b_1$ | 0 | 0 | $d$ | 0 | $k_2$ | 0 | $k_2\alpha_2 + \alpha_0$ |
| chain | $a$ | $a$ | $b_1$ | 0 | $d$ | $d$ | $k_1$ | 0 | $-k_1 b_1$ | $k_1\alpha_1 + \alpha_0$ |
| center | $a$ | $a$ | $b_1$ | $b_2$ | 0 | 0 | $k_1$ | $k_2$ | $-k_1 b_1 - k_2 b_2$ | $k_1\alpha_1 + k_2\alpha_2 + \alpha_0$ |

**Table 3.** Values of the joint probability density, the mean squared displacement and the mean squared velocity for our topology, the center topology, the ring topology, and chain topology in the limits of $t \ll \tau$, $t \gg \tau$ and for $\tau = 0$. Here, the TM, the JPD, the MSD, and the MSV denote the time domain, the joint probability density, the mean squared displacement, and the mean squared velocity, respectively.

| TM | $x, v$ | JPD / MSD, MSV | Topology | | |
|---|---|---|---|---|---|
| | | | Our topology, Center topology | Ring topology | Chain topology |
| $t \ll \tau$ | $x$ | JPD | $\exp[-\frac{2\beta\tau x^2}{\alpha t^2}]$ | $\exp[-\frac{2\beta\tau x^2}{\alpha t^2}]$ | $\exp[-\frac{2\beta\tau x^2}{\alpha t^2}]$ |
| | | MSD | $\frac{\alpha}{4\beta\tau}t^2$ | $\frac{(k_2\alpha_2 + \alpha_0)}{4\beta\tau}t^2$ | $\frac{(k_1\alpha_1 + \alpha_0)}{4\beta\tau}t^2$ |
| | $v$ | JPD | $\exp[-\frac{2\tau v^2}{\alpha(e+r)t^4}]$ | $\exp[-\frac{2\tau v^2}{\alpha(e+r)t^4}]$ | $\exp[-\frac{2\tau v^2}{\alpha(e+r)t^4}]$ |
| | | MSV | $\frac{\alpha(e+r)}{4\tau}t^4$ | $\frac{(k_2\alpha_2 + \alpha_0)}{4\tau}t^4$ | $\frac{(k_1\alpha_1 + \alpha_0)(r - k_1 b_1)}{4\tau}t^4$ |
| $t \gg \tau$ | $x$ | JPD | $\exp[-\frac{x^2}{2\alpha t}]$ | $\exp[-\frac{x^2}{(k_2\alpha_2 + \alpha_0)t}]$ | $\exp[-\frac{x^2}{(k_1\alpha_1 + \alpha_0)t}]$ |
| | | MSD | $\alpha t$ | $(k_2\alpha_2 + \alpha_0)t$ | $(k_1\alpha_1 + \alpha_0)t$ |
| | $v$ | JPD | $\exp[-\frac{v^2}{4\alpha(e+r)^2 t^3}]$ | $\exp[-\frac{v^2}{4(k_2\alpha_2 + \alpha_0)r^2 t^3}]$ | $\exp[-\frac{v^2}{4(k_1\alpha_1 + \alpha_0)(r - k_1 b_1)^2 t^3}]$ |
| | | MSV | $\alpha(e+r)^2 t^3$ | $(k_2\alpha_2 + \alpha_0)rt^4$ | $(k_1\alpha_1 + \alpha_0)(r - k_1 b_1)t^4$ |

| | | | | |
|---|---|---|---|---|
| $\tau=0$ | $x$ | JPD | $\exp[-\frac{2x^2}{\alpha(e+r)t^4}]$ | $\exp[-\frac{2x^2}{\alpha(e+r)t^4}]$ | $\exp[-\frac{2x^2}{\alpha(e+r)t^4}]$ |
| | | MSD | $\alpha t$ | $(k_2\alpha_2+\alpha_0)rt^4$ | $(k_1\alpha_1+\alpha_0)(r-k_1b_1)t^4$ |
| | $v$ | JPD | $\exp[-\frac{v^2}{2\alpha t}]$ | $\exp[-\frac{v^2}{2\alpha t}]$ | $\exp[-\frac{v^2}{2\alpha t}]$ |
| | | MSV | $\alpha t$ | $(k_2\alpha_2+\alpha_0)t$ | $(k_1\alpha_1+\alpha_0)t$ |

## 4. Conclusion

In conclusion, we have derived our equation as the Fokker-Planck equation, subject to an exponentially correlated Gaussian force. We approximately have found the solution of the joint probability density by using double Fourier transforms.

The values of the coupling constants in our topology, the ring topology, the chain topology, and the center topology are summarized in Table 2. The solutions of the Fokker-Planck equation for our topology and the center topology get the same result, because the two topologies have same values of the coupling constant, as seen in Table 3. The mean squared velocity and the mean squared displacement in our topology behavior as the super-diffusion proportional to $t^4$ in $t \ll \tau$ and for $\tau=0$, while the mean squared displacement (the mean squared velocity) has the Gaussian form, which is the normal diffusion proportional to the time $t$ in $t \gg \tau$ ($\tau=0$). We obtain that the moment $\mu_{2,2}$ is proportional to $t^5$ in $t \ll \tau$ and for $\tau=0$, and $t^4$ in $t \gg \tau$, consistent with our result. Other values $\mu_{m,n}$ compared to the high moments will be published elsewhere.

The approximate solution of the Fokker-Planck equation has been solved, which is a simple approximate solution from our model for the first time. We will extend to our model to the generalized Langevin equation or the force equation of motion with other forces [40-46]. The results will be compared and analyzed with the other theories, the computer-simulations, and the experiments. More detailed results for the probability density will be continuously published in the other journals.

**Appendix A: Derivation of** $U(\zeta,v,t)$

In Appendix A, we will derive, Eq. (7), the Fourier transform of the Fokker-Planck equation. First of all, the equations of motion for an active particle subjected to the harmonic and viscous forces, in contact with two heat reservoirs is represented in terms of

$$m_0 \frac{dv(t)}{dt} = -rv(t) - \beta x(t) + k_1 z_1(t) + k_2 z_2(t) + g_0(t), \quad (1)$$

$$m_1 \frac{dz_1(t)}{dt} = -a_1 z_1(t) + b_1 v(t) + d_1 z_2(t) + g_1(t), \quad (2)$$

$$m_2 \frac{dz_2(t)}{dt} = -a_2 z_2(t) + b_2 v(t) + d_2 z_1(t) + g_2(t). \quad (3)$$

In the above equations, we deal with $a_1=a_2, d_1=d_2$ in order to simplify the model. The dimensionless mass is $m_i=1$ for $i=0,1,2$. The Laplace transform of $x(t)$ is defined by $Lx(t)=x(s)$, and the inverse Laplace transform of $x(s)$ is given by $L^{-1}x(s)=x(t)$. After the Laplace transforming of Eqs. (1)-(3), we insert $z_1(s)$ and $z_2(s)$ into the Laplace transform of Eq. (1). Thus, the equation is derived as

$$sv(s) = -rv(s) - \beta x(s) + g_0(s) + [\frac{b_1(s+a)}{(s+a)^2-d^2} + \frac{b_2 d}{(s+a)^2-d^2}]k_1 v(s)$$

$$+[\frac{b_1(s+a)}{(s+a)^2-d^2} + \frac{b_2 d}{(s+a)^2-d^2}]k_2 v(s) + \frac{k_1(s+a)}{(s+a)^2-d^2}g_1(s) + \frac{k_1 d}{(s+a)^2-d^2}g_2(s) + \frac{k_2(s+a)}{(s+a)^2-d^2}g_2(s) + \frac{k_2 d}{(s+a)^2-d^2}g_1(s). \quad \text{(A-1)}$$

The inverse Laplace transform of the above equation is given by

$$\frac{dv(t)}{dt} = -rv(t) - \beta x(t) + g_0(t) + L^{-1}[\frac{b_1(s+a)k_1 v(s)}{(s+a)^2-d^2}] + L^{-1}[\frac{b_2 d k_1 v(s)}{(s+a)^2-d^2}] + L^{-1}[\frac{b_1(s+a)k_2 v(s)}{(s+a)^2-d^2}] + L^{-1}[\frac{b_2 d k_2 v(s)}{(s+a)^2-d^2}] + L^{-1}[\frac{k_1(s+a)}{(s+a)^2-d^2}g_1(s)] + L^{-1}[\frac{k_1 d}{(s+a)^2-d^2}g_2(s)]$$

$$+L^{-1}[\frac{k_2(s+a)}{(s+a)^2-d^2}g_2(s)] + L^{-1}[\frac{k_2 d}{(s+a)^2-d^2}g_1(s)]$$

$$= ①+②+③+④+⑤+⑥+⑦+⑧+⑨+⑩+⑪ \text{ terms.} \quad \text{(A-2)}$$

Inserting the ① and ② terms of Eq. (A-2) into Eq. (4), we calculate as

$$-\frac{\partial}{\partial v}<[①+②]\delta(x-x(t))\delta(v-v(t))> = [-v\frac{\partial}{\partial x}++\beta x\frac{\partial}{\partial v}]<\delta(x-x(t))\delta(v-v(t))> = [-v\frac{\partial}{\partial x}++\beta x\frac{\partial}{\partial v}]U(x,v,t). \quad \text{(A-3)}$$

Inserting the ④, ⑤, ⑥, and ⑦ terms of Eq. (A-2) into Eq. (4), we calculate as

$$-\frac{\partial}{\partial v}<[④+⑤+⑥+⑦]\delta(x-x(t))\delta(v-v(t))> = \frac{\partial}{\partial v}<[(-b_1k_1-b_2k_2)v]\delta(x-x(t))\delta(v-v(t))> = e\frac{\partial}{\partial v}vU(x,v,t) \cdot \quad (A-4)$$

Inserting the ③, ⑧, ⑨, ⑩, and ⑪ terms of Eq. (A-2) into Eq. (4), we lastly calculate as

$$-\frac{\partial}{\partial v}<[③+⑧+⑨+⑩+⑪]\delta(x-x(t))\delta(v-v(t))> = -[\alpha_0+\alpha_1k_1+\alpha_2k_2][C(t)\frac{\partial^2}{\partial x\partial v}-B(t)\frac{\partial^2}{\partial v^2}]<\delta(x-x(t))\delta(v-v(t))>$$

$$= -\alpha[C(t)\frac{\partial^2}{\partial x\partial v}-B(t)\frac{\partial^2}{\partial v^2}]U(x,v,t) \cdot \quad (A-5)$$

Thus, the Fokker-Planck equation, Eq. (5), is satisfied from Eqs. (A-3)-(A-5). We consequently get Eq. (7) after taking the double Fourier transform of the Fokker-Planck equation.